# First-principles Investigations on Polytypes of BaTiO$_3$: Hybrid Calculations and Pressure Dependences

Yu-Seong Seo and Jai Seok Ahn[*]

*Department of Physics, Pusan National University, Busan 609-735, Republic of Korea*

We report our first-principles investigations on three polytypes of BaTiO$_3$ (BTO): a paraelectric phase with cubic *Pm*-3*m* structure and two ferroelectric (FE) phases with tetragonal *P*4*mm* and rhombohedral *R*3*m* structures. We compared the structural and the electrical properties of BTO obtained by using various approaches: e.g., the Hartree-Fock (HF) theory, the density functional theory (DFT) with the local density approximation (LDA) or with the two generalized gradient approximations (two GGAs: PWGGA and PBE), and three hybrid functionals of the HF and the DFT (B3LYP, B3PW, and PBE0). For the *P*4*mm* structure, the two GGAs and the hybrid functionals reproduced the cell volumes, but slightly overestimated the *c*/*a* ratio. The hybrid functionals provided accurate predictions for the experimental energy gaps, but slightly underestimated the experimental dielectric constants. The calculated dielectric constants were inversely proportional to the *c*/*a* ratios for the *P*4*mm* structure (or the $c_H/a_H$ ratio for the *R*3*m* structure), irrespective of the functional choice. Also, the over-estimated polarization could be ascribed to a super-tetragonality in the GGA/hybrid functionals. The pressure dependences for the cell parameters, fractional atomic displacements, energy gaps, dielectric constants, and FE polarizations were calculated by using the B3PW hybrid functional. As pressure was increased, the polarization decreased monotonically until it reached zero at a critical pressure of ~ 20 GPa for both the *P*4*mm* and the *R*3*m* structures. Anomalous behaviors were also observed in the atomic movements and the polarizations for the *P*4*mm* structure: $\delta_{Ti}$, the fractional coordinate of Ti, showed a sign reversal at pressures below −8 GPa while the polarization showed a maximum at −2 GPa and then decreased with decreasing pressure. However, such effects were not observed for the *R*3*m* structure: the polarization monotonically increased with decreasing pressure. Such behaviors of the polarizations, together with super-tetragonality/super-trigonality, are discussed.

* Email: jaisahn@pusan.ac.kr

 Fax: +82-51-513-7664 (Dept.)

## I. INTRODUCTION

Ferroelectric (FE) materials have a wide range of applications, such as information storage, energy conversion, and medical imaging [1, 2]. A FE material has a large FE spontaneous polarization and a large piezoelectric coefficient, $d_{33}$, because it belongs to a subgroup of the non-centosymmetric piezoelectric group [2]. The phase diagram of oxide FE perovskites, such as Pb(Zr,Ti)O$_3$ (PZT), shows three phases, rhombohedral, tetragonal and cubic, connected at a triple point. Ferroelectricity occurs at the two phases, tetragonal and rhombohedral, that are connected by a morphotropic phase boundary (MPB) composed of intermediate monoclinic/orthorhombic phases [3-7]. The existence of intermediate phases is believed to be indispensible for a polarization rotation and for a large $d_{33}$ [8], such as in PZT or Ba(Ti,Zr)O$_3$-$x$(Ba,Ca)TiO$_3$ (BTZ-$x$BCT) across the MPB [7, 9]. Recently, we investigated the relaxor-ferroelectrics BTZ-$x$BCT and found that the local structure, the polar nano regions, and the dipolar domain-dynamics are also important ingredients to understand their properties [10-12]. A neutron pair-distribution-function (PDF) analysis provided notable local structural features departing from the given crystallographic structures, such as local Ti distortions toward the <111> direction, a trigonal 3:3 Ti-O length distribution, and a single Zr-O bond length in BTZ-$x$BCT [10, 11]. A Raman spectroscopy on BTZ-$x$BCT showed a rich interplay between the long-range average structure and the short-range local orders, with that interplay occurring as a function of pressure [12].

FE BaTiO$_3$ (BTO) has been investigated using various first-principles methods [13-18]. Recently, Wu and Cohen proposed a new PBE ansatz that more accurately predicted the tetragonality factor [19]. The effects of pressure on the properties of BTO, however, have been rarely studied with first-principle methods until now. Results are only available for the tetragonal phase calculated with the local density approximation (LDA) [18]. To further understand the polytypes of BTO under different pressure, we employed first-principles calculations using hybrid functionals [20], that were composites of the Hartree-Fock (HF) theory and the Kohn-Sham density functional theory (DFT). Among the approximations to the DFT, the LDA is well known to underestimate the lattice constants while the generalized gradient approximation (GGA) [21] provides an improved prediction for the equilibrium volume over the LDA. The LDA and the GGA are well known to significantly underestimate the band gaps in terms of the Kohn-Sham/quasi-particle energy bands. On the other hand, the HF method overestimates the band gaps. Thus, so-called "hybrid" functionals that combine the HF theory and the DFT, such as B3LYP [22], B3PW [23], and PBE0 [24], are very attractive as they can provide accurate values of the bond lengths, energy band structures, and phonon frequencies [15]. In this manuscript, we consider three polytypes of BTO: a paraelectric phase with cubic *Pm*-3*m* structure and two FE phases with tetragonal *P*4*mm* and rhombohedral *R*3*m* structures. We compared the structural and the electrical properties obtained by using various functionals and chose the B3PW functional for subsequent calculations to determine the pressure dependences of the properties of BTO.

## II. COMPUTAIONAL DETAILS

First-principles calculations were performed with the HF, DFT, and hybrid methods as implemented in the CRYSTAL09 code [22]. We used basis sets (BS) composed of Gaussian-type orbitals. To describe heavy elements, such as barium and titanium, we adopted the Hay-Wadt effective core potentials with small cores for the BS's. Ba ($5s/5p/6s$) and Ti ($3s/3p/4s/3d$) electrons were considered as valence electrons for self-consistent calculations [25]. Thus, the computational time could be minimized within our computational resources. For the oxygen atom, the full electron BS O-411d11G was used [26]. The reciprocal space integration was approximated by sampling the Brillouin zone with a 6×6×6 mesh of the

Monkhorst-Pack scheme. The self-consistent cycles were repeated until a convergence was reached within a tolerance of $10^{-10}$ Ha (~ $3\times10^{-9}$ eV) for the total energy difference.

We compared the ground-state configurations calculated with a number of different exchange-correlation functionals from among those of density functional theory (DFT), the Hartree-Fock (HF) approximation, and their hybrids. This comparison allowed us to select the exchange-correlation functional, that best described the lattice-structural and the electronic properties of the three different structures of BTO. For the DFT, the LDA and two different-forms of GGAs (PBE [21] and PWGGA [27]) were used. Usually, the HF method overestimates the experimental gap; on the other hand, the Kohn-Sham energy gap calculated in the DFT usually underestimates it. Therefore, by applying hybrids of the two methods, one can obtain a more accurate band gap. We applied three functional forms for hybrid methods: B3LYP [22], B3PW [23], and PBE0 [24]. The B3LYP has a Becke's three-parameter hybrid functional form [23], which includes the HF exchange and the non-local correlation of Lee *et al*. [28], with *A* = 0.2, *B* = 0.9, and *C* = 0.81:

$$E_{xc} = (1 - A) \cdot (E_x^{LDA} + B \cdot E_x^{Becke}) + A \cdot E_x^{HF} + (1 - C) \cdot E_c^{VWN} + C \cdot E_c^{LYP}, \qquad (1)$$

where $E_x^{LDA}$ and $E_c^{VWN}$ are, respectively, the LDA exchange and the Vosko-Wilk-Nusair correlation [29] that fits the Ceperley-Alder data for an electron gas [30]. Similarly, The B3PW hybrid functional uses the PWGGA, $E_c^{PWGGA}$, as a non-local correlation instead of the $E_c^{LYP}$ used for the B3LYP shown in Eq. (1). The PBE0 is a parameter-free hybrid functional, which uses a PBE (GGA) exchange-correlation mixed with a 25% HF exchange.

The equilibrium geometries were found by optimizing the lattice constants and the fractional displacements of the basis atoms for the cubic (*Pm-3m*), tetragonal (*P4mm*), and rhombohedral (*R3m*) structures of BTO: structural relaxations were repeated until the norm of the total force became less than $3\times10^{-5}$ Ha/Bohr ($\approx 1.5\times10^{-3}$ eV/Å ). High-frequency (optical) dielectric constants were calculated with the finite field perturbation method: the effect of a "sawtooth" electric potential applied to a supercell was evaluated numerically. For the tetragonal or the rhombohedral (in hexagonal setting) case, the anisotropic dielectric constants, $\varepsilon_{a,\infty}$ and $\varepsilon_{c,\infty}$, were determined by applying an electric field to a supercell along the crystallographic *a* and *c* directions, respectively. The spontaneous polarization of the FE phases were evaluated through the Berry phase approach [31] by comparing the stable non-centosymmetric phase (with geometric parameter $\lambda = 1$) to the unstable centosymmetric phase (with $\lambda = 0$). Geometry optimization was fulfilled under a hydrostatic pressure for further pressure-dependent calculations.

## III. RESULTS AND DISCUSSION

**1. Structural and Electronic Properties: Comparison of HF, DFT, and Hybrid Methods**

To investigate the three polytypes of BTO, we constrained the symmetries as follows: the paraelectric cubic (*C*) *Pm*-3*m*, the ferroelectric tetragonal (*T*) *P*4*mm* and rhombohedral (*R*) *R*3*m*. Fully-relaxed results obtained with seven exchange-correlation functionals—HF, LDA, PWGGA, PBE(GGA), B3LYP, B3PW, and PBE0— are summarized in Table 1.

As expected, the LDA underestimated the experimental lattice constants, and the other functionals gave better, but over-estimated, results. Moreover, all other functionals except the LDA overestimated the tetragonality factor, defined as *c*/*a*, in *P*4*mm*. These are the well-known limitations of GGA functionals other than the WC-GGA functional [19]. Among the hybrid functionals, the B3LYP largely overestimated the volume and the tetragonality, and similar behaviors were found from the reports of other group who used the same functional [15].

The DFT functionals, LDA and GGAs, underestimated the experimental energy gap, $E_g$, and the HF gave ~12 eV, a 3-4 times overestimate, that overestimate being an already well-known result. The hybrid functionals provided very close estimates, and among the hybrid functional, the B3LYP and the B3PW functional gave better predictions than the PBE0 functional. For the $R3m$ structure, an experimental $E_g$ is unavailable to our knowledge, so we compared the $E_g$ to a value of 4.6 ± 0.1 eV obtained from amorphous BTO [32], which should be useful as a maximum bound for the experimental value of $E_g$ for the $R3m$ structure.

The electronic dielectric constant $\varepsilon_\infty$ was found to be substantially lower than the experimental data when the HF was used, while it was overestimated when the DFT functionals were used. The overestimates by the DFT were well-interpreted as being due to the missing polarization-dependence in the exchange-correlation functionals [33], such as a semi-empirical correction for the dispersion interaction in Grimme scheme [34]. On the other hand, hybrid functionals provided in-between, but slightly underestimated, values. The anisotropic dielectric constants, $\varepsilon_{a,\infty}$ and $\varepsilon_{c,\infty}$, also showed similar behaviors. We carefully compare the values, $\varepsilon_{a,\infty}$'s and $\varepsilon_{c,\infty}$'s, in Table 1 to find possible interrelationships. We found two simple relationships between them. Calculated $\varepsilon_{c,\infty}$'s are found to be almost proportional to $\varepsilon_{a,\infty}$'s. In addition, more intriguingly, the values of $\varepsilon_{a,\infty}$ or $\varepsilon_{c,\infty}$ are inversely proportional to the values of $c/a$ (or $c_H/a_H$ in the $R3m$ structure).

The polarizations, $P_s$'s, of the FE phases were calculated using the Berry phase method with seven different exchange-correlation functionals and they are also summarized in Table 1. The LDA underestimated the experimental data while the other functionals overestimated these data. The underestimates of LDA are due to an underreckoning of the cell volume because polarization is proportional to the degree of lattice distortion along the [001]/[111] direction in the $P4mm/R3m$ structures. We used the HF pseudo-potential for Ti, which is known to give an overestimated polarization. The pseudo-potential is known to give a 20-30% overestimate of the polarization compared with the all-electron BS. A better estimate can be obtained if one uses optimized basis sets to remove super-tetragonality by tuning $c/a$ ratio[15].

## 2. Pressure Dependence

First, we compare the pressure dependences of the optimized structures of the tetragonal and the rhombohedral phases of BTO. The *ab-initio* calculations were done with the B3PW hybrid functional. The lattice constants and the fractional atomic displacements were fully relaxed under external hydrostatic pressures ranging from −10 to 20 GPa. The calculated cell volumes of the BTO polytypes, *Pm*-3*m* cubic (○), *P*4*mm* tetragonal (□), and *R*3*m* rhombohedral (◊) structures, are shown in Fig. 1(a) as functions of pressure. As the pressure increases, their volumes decrease monotonically, and the volume of the *P*4*mm* structure is close to that of the *R*3*m* structure at pressures from -2 to 20 GPa while the volume difference becomes evident at pressures below −8 GPa. The pressure dependences of the tetragonality, the $c/a$ ratio of *P*4*mm* structure, and the rhombohedral angle $\alpha$ of the *R*3*m* structure are also shown in Fig. 1(b). The tetragonality becomes larger when a tensile stress is applied, and it increases abruptly at pressures below −6 GPa. When a compressive stress is applied, the $c/a$ ratio decreases to an experimental value of 1.010 [38] at 8-10 GPa and eventually converges to one. As mentioned earlier, the super-tetragonality is a well-known limitation of GGA-derived functionals when optimized BS's are not used [15]. The rhombohedral angle $\alpha$ decreases rapidly when negative pressure is applied while it increases to an experimental value of 89.84$^\circ$ [41] at 4-6 GPa and further increases gradually to 90$^\circ$ as more positive pressure is applied.

Figures 2(a) and 2(b) show the fractional atomic displacements for the *P*4*mm* and the *R*3*m* structures fully relaxed under external hydrostatic pressure. In tetragonal BTO with the

$P4mm$ structure, basis atoms are given in Wyckoff positions and asymmetric coordinates as Ba 1a:(0, 0, 0), Ti 1b:(1/2, 1/2, 1/2 + $\delta_{Ti}$), O$_1$ 1b:(1/2, 1/2, $\delta_{O1}$), and O$_2$ 2c:(1/2, 0, 1/2 + $\delta_{O2}$). The basis atoms of the rhombohedral BTO with the $R3m$ structure are located at Ba 1a:(0, 0, 0), Ti 1a:(1/2 + $\delta_{Ti}$, 1/2 + $\delta_{Ti}$, 1/2 + $\delta_{Ti}$), and O 3b:(1/2 + $\delta_{Oc}$, 1/2 + $\delta_{Oc}$, $\delta_{Oa}$ + $\delta_{Oc}$). Here, 1a, 1b, 2c, and 3b are the Wyckoff positions of the atoms. The fractional atomic displacements, $\delta_{Ti}$, $\delta_{O1}$, and $\delta_{O2}$, of the tetragonal phase are shown in Fig. 2(a) as functions of pressure. They show atomic movements along the [001] direction. Titanium goes to [001] while oxygen goes to [00-1] for pressures ranging from −6 to 20 GPa. The $\delta_{Ti}$, however, changes its sign: it becomes negative at pressures below −8 GPa. Apical oxygen O$_1$ and planar oxygen O$_2$, forming a dipole moment with titanium, move much faster than titanium. Thus, the polarization is maintained along the [001] direction for all pressure values, even when there is a sign reversal in $\delta_{Ti}$. The $\delta_{Ti}$ reaches its maximum value of ~ 0.02 at a pressure between −2 and 0 GPa. The pressure dependences of $\delta_{Ti}$, $\delta_{Oa}$, and $\delta_{Oc}$ of the rhombohedral phase are also shown in Fig. 2(b). They show atomic movements along the [111] direction mixed with the [001] direction. Titanium goes to the [111] direction while oxygen goes to the [-1-1-1] and the [00-1] directions monotonically for all pressure values. Thus, the polarization is maintained along the [111] direction, and the absolute values of $\delta_{Ti}$, $\delta_{Oa}$, and $\delta_{Oc}$ increase as the pressure is decreased.

The pressure dependences of the Kohn-Sham (mixed with HF) energy gaps $E_g$'s of the $Pm$-$3m$, $P4mm$, and $R3m$ phases of BTO are shown in Fig. 3(a). The cubic and the tetragonal phases show almost monotonic increases of $E_g$'s while $E_g$ calculated for the rhombohedral phase decreases as the pressure is increased. Hybridization between orbitals should be more active when the pressure is increased; thus, $E_g$ should increase with increasing pressure as predicted for the cubic and the tetragonal phases. Therefore, the decreasing $E_g$ in the rhombohedral phase is a counter-intuitive behavior, but experimental data support its values at least at zero pressure. Note the experimental data at ambient pressure, which are shown with filled symbols in Fig. 3(a). Moreover, as the pressure is increased, the rhombohedral structure collapses into the cubic structure eventually. The larger $E_g$ of the $R3m$ structure decreases as a function of pressure and it coincides with the value of the cubic structure, as shown in Fig. 3(a).

The optical dielectric constants $\varepsilon_\infty$ were calculated for the $Pm$-$3m$, $P4mm$, and $R3m$ phases as functions of pressure, as shown in Fig. 3(b). The calculated $\varepsilon_\infty$ of the $Pm$-$3m$ phase increases with increasing pressure, but its pressure dependence is very gentle: $\varepsilon_\infty(p) \approx 4.94 + 0.0022 \times p$[GPa]. On the other hand, the two ferroelectric phases show strong pressure dependences of the dielectric constants, $\varepsilon_{a,\infty}$'s and $\varepsilon_{c,\infty}$'s, irrespective of the directions of the electric field. The $\varepsilon_{c,\infty}$'s, the dielectric constants along the symmetry axis, are smaller than the $\varepsilon_{a,\infty}$'s, and the difference between the two decreases as a function of pressure and should disappear eventually, such that the two merge to the value of the cubic phase at high pressures. The $\varepsilon_{a,\infty}$ and $\varepsilon_{c,\infty}$ of the $R3m$ phase, however, converge to ~ 5.2, which is unambiguously higher than the converged value of 4.9/5.0 of the $Pm$-$3m$/$P4mm$ phase. Moreover, no field dependence is found in the above behavior for $E$-field variations of $2.6 \times 10^6$ - $2.6 \times 10^9$ V/m. Thus, the above-mentioned small difference in the converged values of the dielectric constants can be ascribed to an unpredicted limitation of finite-field perturbation methods.

The spontaneous polarizations, $P_s$'s, of the ferroelectric structures, $P4mm$ and $R3m$, were also calculated as functions of pressure, as shown in Fig. 4. The polarization vectors were parallel to the [001] ($P4mm$) and the [111] ($R3m$) directions of the pristine cubic lattice. Note that the calculated polarizations in Fig. 4 overestimate the experimental values at ambient pressure. As mentioned above, $c/a$ ratio recovers its experimental value at 8-10 GPa in the $P4mm$ structure, such that $P_s$ becomes ~ 25.7 $\mu$C/cm$^2$, close to the reported value of 26.0 $\mu$C/cm$^2$ [40]. The same argument can be applied to the rhombohedral system using a trigonality $c_H/a_H$ ratio or angle $\alpha$. The angle $\alpha$ regains its experimental value at 4-6 GPa in the $R3m$ structure, and $P_s$ becomes ~ 37.7 $\mu$C/cm$^2$, which is close to the reported value of 34 $\mu$C/cm$^2$ [43]. Therefore, the overestimated $P_s$'s can be unambiguously ascribed to the super-tetragonality/super-trigonality inherent in calculations using GGA-derived functionals. As the pressure is increased, $P_s$ decreases monotonically until it is reduced to zero at a critical pressure of ~ 20 GPa for both the $P4mm$ and the $R3m$ structures. This diminishing $P_s$ as a function of pressure is consistent with the general tendency of pressure-induced reductions in the lattice distortions. The critical pressure, however, overestimates the experimental observations: the $T$-to-$C$ transition is at 3.4 GPa at room temperature and the tricritical point is expected at 6.5 GPa [44, 45]. If we consider the above-mentioned overestimation factor due to the super-tetragonality/super-trigonality when using the GGAs for our calculations, then we can obtain better predictions of $P_s$. When a negative pressure is applied, $P_s$ increases monotonically in the $R3m$ structure while it reaches a maximum value of 49 $\mu$C/cm$^2$ at ~ −2 GPa and then decreases in the $P4mm$ structure. In our calculations for the $P4mm$ structure, tetragonality increases gradually with decreasing pressure, and then increases abruptly at pressures below ~ −6 GPa, as shown in Fig. 1(b), while $P_s$ decreases at pressures below ~ −4 GPa. These predictions are partly consistent with other reports on PbTiO$_3$ and BaTiO$_3$ using the LDA functional. The continuous increase of $P_s$ with decreasing pressure at negative pressures has been predicted in PbTiO$_3$: it was ascribed to an increasing tetragonality [46]. Also, a calculation for an epitaxial (tetragonal) strain on BaTiO$_3$ predicted an increasing $P_s$ and did not show any decreasing behavior of $P_s$ below a certain negative pressure [47].

## IV. CONCLUSION

We calculated the structural and the electronic properties of three polytypes of BaTiO$_3$, *Pm*-3*m*, *P*4*mm*, and *R*3*m* structures, by using *ab-initio* calculations with DFT, HF, and hybrid functionals. Results obtained for fully-relaxed geometries by using seven different functionals were compared. Lattice parameters, energy gaps, dielectric constants, and FE polarizations were calculated for the three polytypes. Pressure dependences were calculated with the B3PW hybrid functional and were compared to previous predictions. Several intriguing and unexpected features were found. The energy gap of the *R*3*m* structure decreases with increasing pressure. The polarization of the *P*4*mm* structure showed an anomalously diminishing behavior with decreasing pressure for negative pressures below −2 GPa, and the fractional displacements of titanium showed a sign reversal at the same time.

## ACKNOWLEDGMENTS

This work was supported by the National Research Foundation of Korea (NRF) grant funded by the Ministry of Education, Science and Technology (MEST, No. 2012006641). The computation is supported by the Korea Institute of Science and Technology Information (KISTI) Supercomputing Center through contract no. KSC-2012-C2-36.

Table 1. Structural and electronic properties of cubic, tetragonal, and rhombohedral phases of BaTiO$_3$ calculated by using HF, DFT, and hybrid functionals. The unit of polarization is $\mu$C/cm$^2$. The last two columns show the experimental data and the data calculated with the LDA. The experimental data were measured at room temperature. The volume for the rhombohedral cell is calculated as $V = a_R^3\sqrt{1 - 3\cos^2\alpha + 2\cos^3\alpha}$.

| Method | HF | LDA | GGA | | HYBRIDS | | | Expt. | Theory |
| | | | PWGGA | PBE | B3LYP | B3PW | PBE0 | | |
| --- | --- | --- | --- | --- | --- | --- | --- | --- | --- |
| *Cubic Pm-3m* | | | | | | | | | |
| $a$ (Å) | 4.008 | 3.936 | 4.004 | 4.008 | 4.016 | 3.980 | 3.970 | 3.996 [35] | 3.958 [15] |
| $V$ (Å$^3$) | 64.40 | 60.97 | 64.20 | 64.37 | 64.75 | 63.06 | 62.55 | 63.81 [35] | 62.01 [15] |
| $E_g$ (eV) | 11.5 | 2.0 | 2.0 | 1.9 | 3.5 | 3.5 | 4.0 | 3.2 [36] | 2.2 [15] |
| $\varepsilon_\infty$ | 3.30 | 5.81 | 5.78 | 5.79 | 4.91 | 4.94 | 4.80 | 5.40 [37] | 6.75 [14] |
| | | | | | | | | | |
| *Tetragonal P4mm* | | | | | | | | | |
| $a$ (Å) | 3.967 | 3.950 | 4.003 | 4.006 | 3.989 | 3.973 | 3.965 | 3.995 [38] | 3.915 [18] |
| $c$ (Å) | 4.271 | 3.984 | 4.154 | 4.164 | 4.275 | 4.157 | 4.131 | 4.034 [38] | 3.995 [18] |
| $c/a$ | 1.077 | 1.009 | 1.038 | 1.039 | 1.072 | 1.046 | 1.042 | 1.010 [38] | 1.020 [18] |
| $V$ (Å$^3$) | 67.22 | 62.15 | 66.56 | 66.84 | 68.03 | 65.61 | 64.94 | 64.36 [38] | 61.23 [18] |
| $E_g$ (eV) | 11.9 | 2.0 | 2.0 | 2.0 | 3.6 | 3.6 | 4.0 | 3.4 [15] | 1.9 [17] |
| $\varepsilon_{a,\infty}$ | 3.10 | 5.71 | 5.43 | 5.42 | 4.47 | 4.61 | 4.51 | 5.19 [39] | 6.20 [17] |
| $\varepsilon_{c,\infty}$ | 3.03 | 5.43 | 4.74 | 4.71 | 3.92 | 4.12 | 4.09 | 5.05 [39] | 5.81 [17] |
| $P_s$ | 48.0 | 22.6 | 40.6 | 41.2 | 48.7 | 44.4 | 42.7 | 26.0 [40] | 28.0 [18] |
| | | | | | | | | | |
| *Rhombohedral R3m* | | | | | | | | | |
| $a_R$ (Å) | 4.042 | 3.962 | 4.058 | 4.065 | 4.092 | 4.037 | 4.023 | 4.004 [41] | 3.966 [17] |
| $\alpha$ (Deg.) | 89.791 | 89.961 | 89.778 | 89.755 | 89.567 | 89.724 | 89.746 | 89.840 [41] | 89.96 [17] |
| $V$ (Å$^3$) | 66.04 | 62.20 | 66.84 | 67.17 | 68.49 | 65.80 | 65.10 | 64.17 [41] | 62.38 [17] |
| $E_g$ (eV) | 12.5 | 2.2 | 2.7 | 2.7 | 4.7 | 4.5 | 4.9 | 4.6 [42] | 2.2 [17] |
| $\varepsilon_{a,\infty}$ | 3.24 | 5.89 | 5.34 | 5.30 | 4.36 | 4.59 | 4.51 | 6.19 [42] | 6.05 [17] |
| $\varepsilon_{c,\infty}$ | 3.05 | 5.69 | 4.80 | 4.75 | 3.73 | 4.10 | 4.06 | 5.88 [42] | 5.82 [17] |
| $P_s$ | 37.0 | 25.6 | 44.9 | 46.1 | 54.2 | 46.9 | 45.2 | 34 [43] | 35.0 [16] |

Figure Captions

Fig. 1. (a) Pressure dependence of unit cell volumes of BaTiO$_3$, calculated with the B3PW hybrid functional. Calculated unit cell volumes are shown with symbols: *Pm-3m* cubic (○), *P4mm* tetragonal (□), and *R3m* rhombohedral (◊). (b) Pressure dependences of the *c*/*a* ratio of *P4mm* (□) and the rhombohedral angle *α* of *R3m* (◊) in degrees.

Fig. 2. Relaxed fractional atomic displacements, (a) $\delta_{Ti}$, $\delta_{O1}$, and $\delta_{O2}$ in *P4mm* and (b) $\delta_{Ti}$, $\delta_{Oa}$, and $\delta_{Oc}$ in *R3m*, as functions of pressure. Note that the fractional coordinates of the basis atoms are as follows: in *P4mm*, Ba (0, 0, 0), Ti (1/2, 1/2, 1/2 + $\delta_{Ti}$), O$_1$ (1/2, 1/2, $\delta_{O1}$), and O$_2$ (1/2, 0, 1/2 + $\delta_{O2}$); in *R3m*, Ba (0, 0, 0), Ti (1/2 + $\delta_{Ti}$, 1/2 + $\delta_{Ti}$, 1/2 + $\delta_{Ti}$), and O (1/2 + $\delta_{Oc}$, 1/2 + $\delta_{Oc}$, $\delta_{Oa}$ + $\delta_{Oc}$).

Fig. 3. Pressure dependence of the (a) Kohn-Sham (mixed with Hartree-Fock) energy gaps and the (b) dielectric constants of *Pm-3m* (○), *P4mm* (□), and *R3m* (◊). Experimental energy gaps are also shown in (a) as filled symbols: *Pm-3m* (●), *P4mm* (■), and *R3m* (◆) from Refs. 36, 15, and 32, respectively.

Fig. 4. Calculated spontaneous polarizations of the ferroelectric structures of BaTiO$_3$, *P4mm* (□) and *R3m* (◊), as functions of pressure. Polarization vectors are parallel to the [001] (*P4mm*) and to the [111] (*R3m*) directions of the pristine cubic lattice.

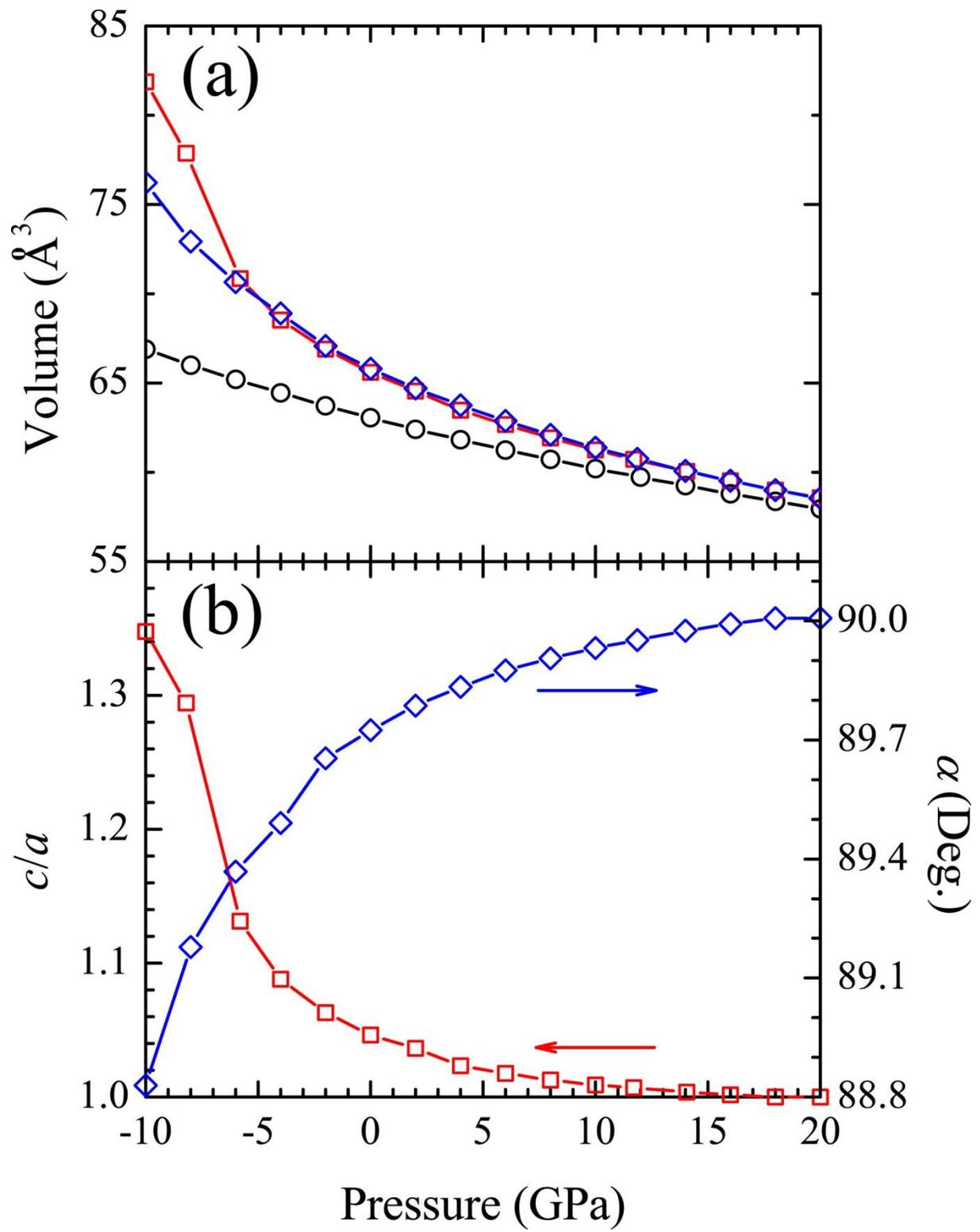

Fig. 1 of Seo et al.

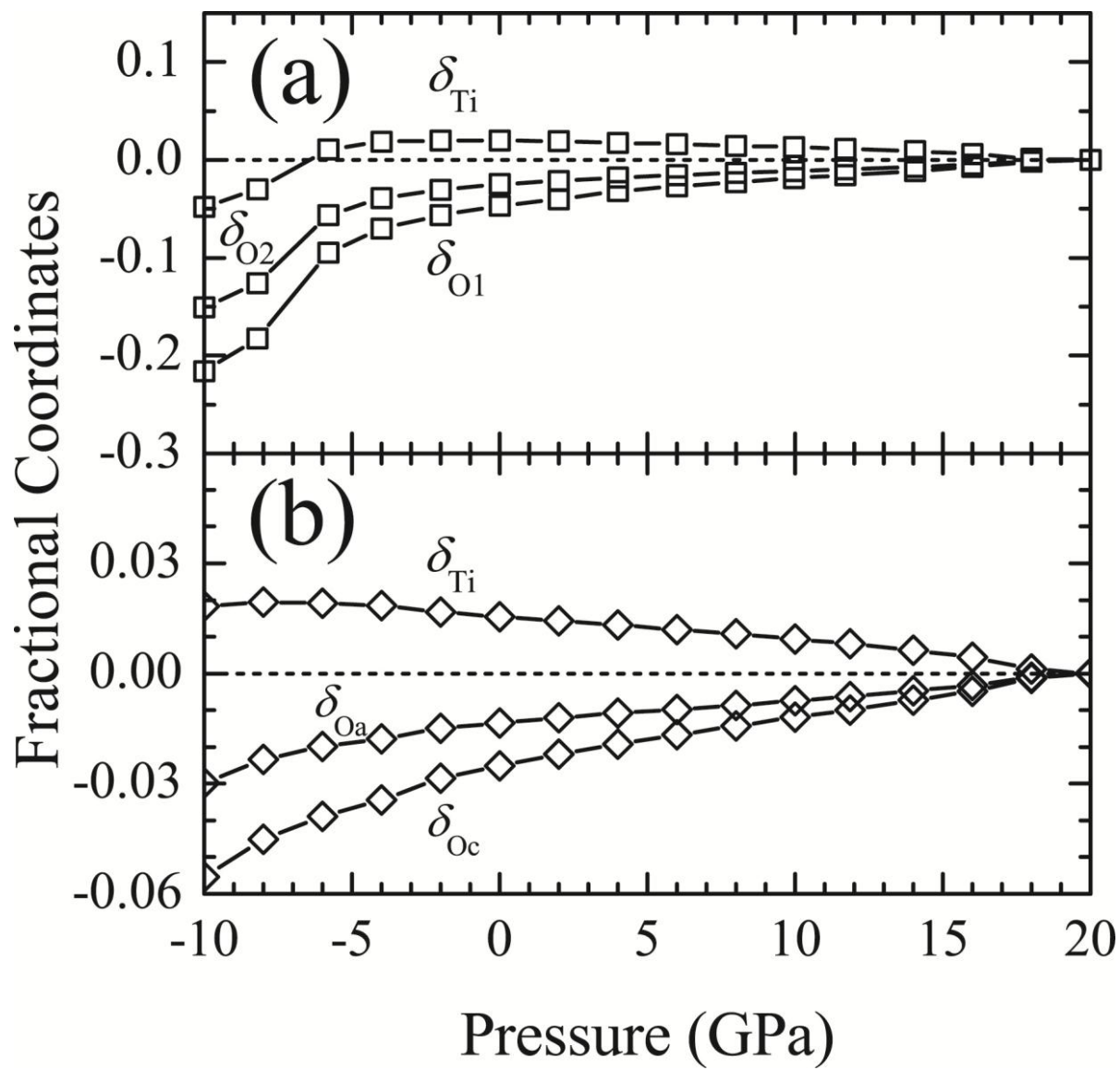

Fig. 2 of Seo et al.

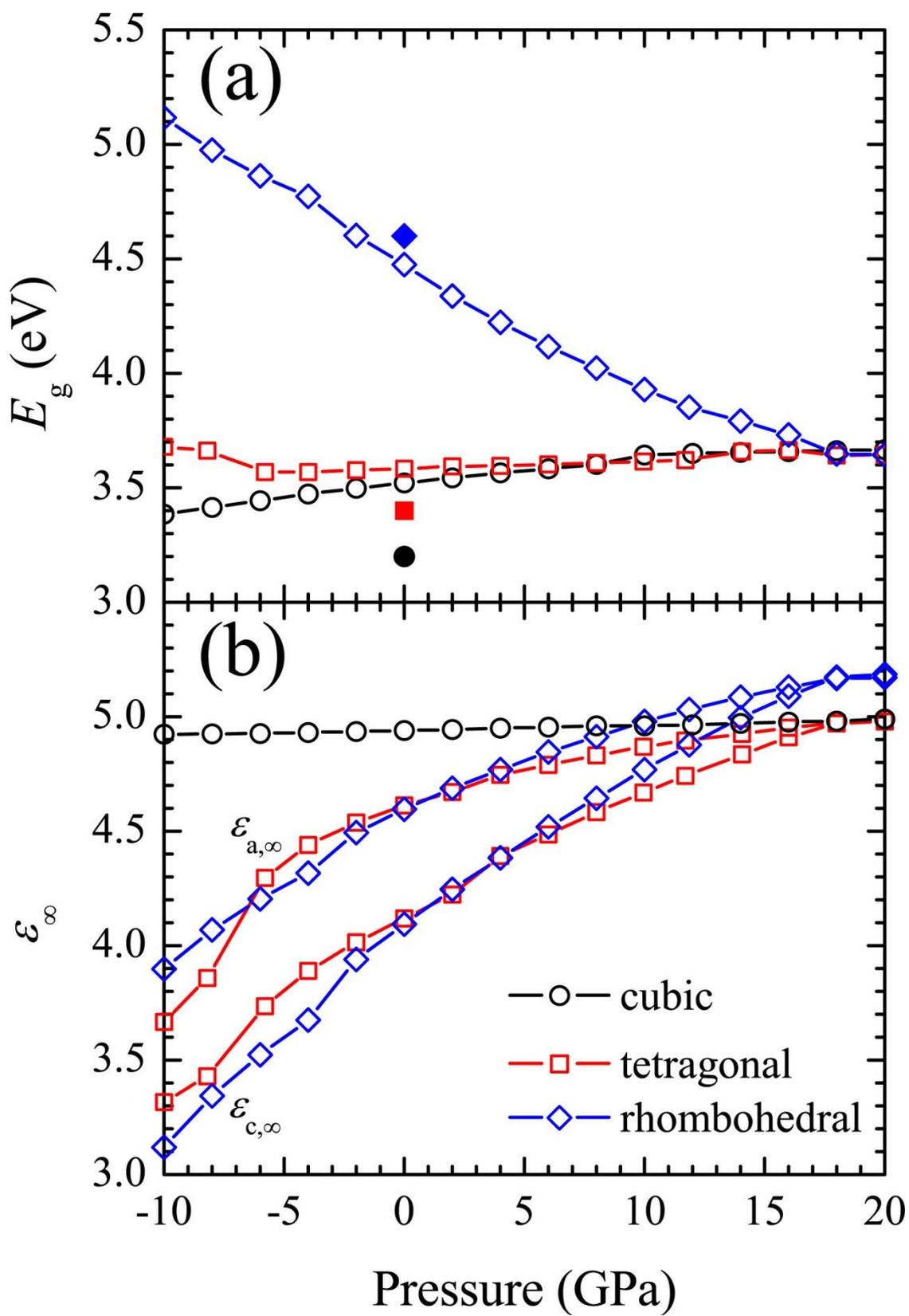

Fig. 3 of Seo et al.

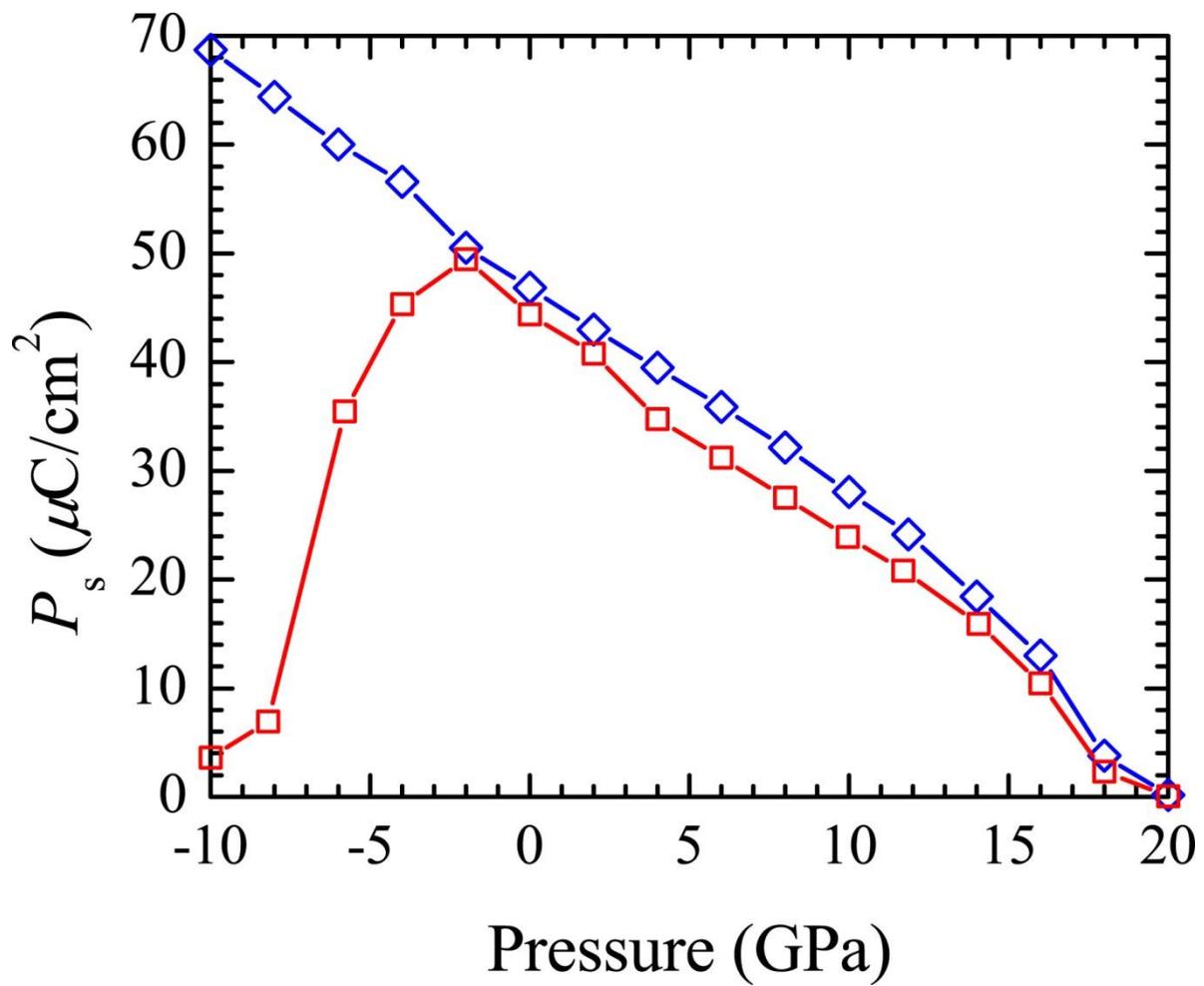

Fig. 4 of Seo et al.